\pdfoutput=1
\documentclass[12pt]{iopart}

\usepackage{iopams}

\usepackage{graphicx}

\usepackage{enumerate}
\usepackage{wasysym}

\usepackage{color}
\usepackage[usenames,dvipsnames]{xcolor}
\usepackage{braket}
\usepackage{siunitx}

\usepackage{multirow}
\begin{document}

\title[]{VUV irradiance measurement of a 2.45 GHz microwave-driven hydrogen discharge}

\author{J. Komppula$^1$, O. Tarvainen$^1$, T. Kalvas$^1$, H. Koivisto$^1$, R. Kronholm$^1$, J. Laulainen$^1$ and P. Myllyperki\"{o}$^2$}

\address{$^1$ University of Jyvaskyla, Department of Physics, \\P.O. Box 35, FI-40014 University of Jyvaskyla, Finland}
\address{$^2$ University of Jyvaskyla, Department of Chemistry, \\P.O. Box 35, FI-40014 University of Jyvaskyla, Finland}
\ead{jani.komppula@jyu.fi}

\vspace{10pt}

\begin{abstract}
Absolute values of VUV-emission of a 2.45 GHz microwave-driven hydrogen discharge are reported. The measurements were performed with a robust and straightforward method based on a photodiode and optical filters. It was found that the volumetric photon emission rate in the VUV-range (80--250~nm) is $10^{16}$--$10^{17}$~1/cm$^3$s, which corresponds to approximately 8\% dissipation of injected microwave power by VUV photon emission. The volumetric emission of characteristic emission bands was utilized to diagnostics of molecular plasma processes including volumetric rates of ionization, dissociation and excitation to high vibrational levels and metastable states. The estimated reaction rates imply that each injected molecule experiences several inelastic electron impact collisions. The upper limit for the total density of metastable neutrals ($2S$ atoms and $c^3\Pi_u$ molecules) was estimated to be approximately 0.5\% of the neutral gas density.
\end{abstract}

%
\vspace{2pc}
%
%
\maketitle
%
\section{Introduction}
Vacuum ultraviolet (VUV) emission of plasmas is rarely applied, but increasingly important tool in applications and diagnostics of plasma sources. VUV-radiation can, for example, significantly affect material processing (\cite{Potts_2013_VUV_ALD,plasma_ALD_review,Tian_2015_Controlling_VUV_in_material_processing} and references therein), and it can be utilized in photolithography for decreasing the length scale of semiconductor components \cite{plasma_etching_review,Stephens_2015_Ly_alpha_optimization}. Furthermore, VUV-emission could play an important role in plasma dynamics via photoelectron emission \cite{Laulainen_NIBS_2014} or cleaning of plasma chamber walls \cite{UV_cleaning}. It could possibly be utilized in development of negative hydrogen ion sources for neutral beam injection into magnetically confined fusion devices and particle accelerators by providing alternatives for the usage of caesium seeding \cite{Tarvainen2011113}. VUV-emission is applied for diagnosing plasma disruptions in thermonuclear fusion devices \cite{fusion_VUV_diagnostic}. In the case of molecular hydrogen plasmas it provides straightforward and robust diagnostics for plasma processes, such as density of hot electrons and volumetric rates of ionization, molecule dissociation and excitation to high vibrational levels \cite{Komppula_2015_VUV_diagnostics,komppula_NIBS_2012}.

This paper describes a study on the VUV-emission of a hydrogen plasma sustained in a microwave discharge ion source. The absolute values of photon irradiance were measured with an SXUV photodiode, which has a uniform response at VUV-wavelengths. The VUV-irradiance in the entire UV-range (100--250~nm) and in emission bands characteristic to hydrogen (110--130~nm, 150--180~nm and 180--220~nm) have been measured by filtering the emission spectrum with UV-FS/BK7 windows or VUV optical bandpass filters. The volumetric photon emission from the line of sight plasma volume was determined for the characteristic emission bands and the information was utilized for diagnostics of plasma process rates as presented in Ref. \cite{Komppula_2015_VUV_diagnostics}. The effects of the ion source operating parameters (pressure, microwave power and magnetic field) on the VUV-emission were compared to those on the extracted ion beam current. 

A similar study has been performed earlier with hydrogen plasma sustained by a filament driven multicusp arc discharge \cite{komppula_NIBS_2012,Komppula_2015_VUV_diagnostics}. There it was found that at least 15--30\% of the discharge power is dissipated via VUV-emission, the VUV-emission is linearly proportional to the discharge power and the VUV-radiation is dominantly emitted by hydrogen molecules. The results measured with the microwave discharge can be compared to the results obtained with the arc discharge. The measurement method is reported thoroughly in this paper since the description of the earlier experiment \cite{komppula_NIBS_2012} was published as a brief and incomplete article in conference proceedings.

\section{Plasma VUV-irradiance measurement}
\begin{figure}
 \begin{center}
 \includegraphics[width=0.70\textwidth]{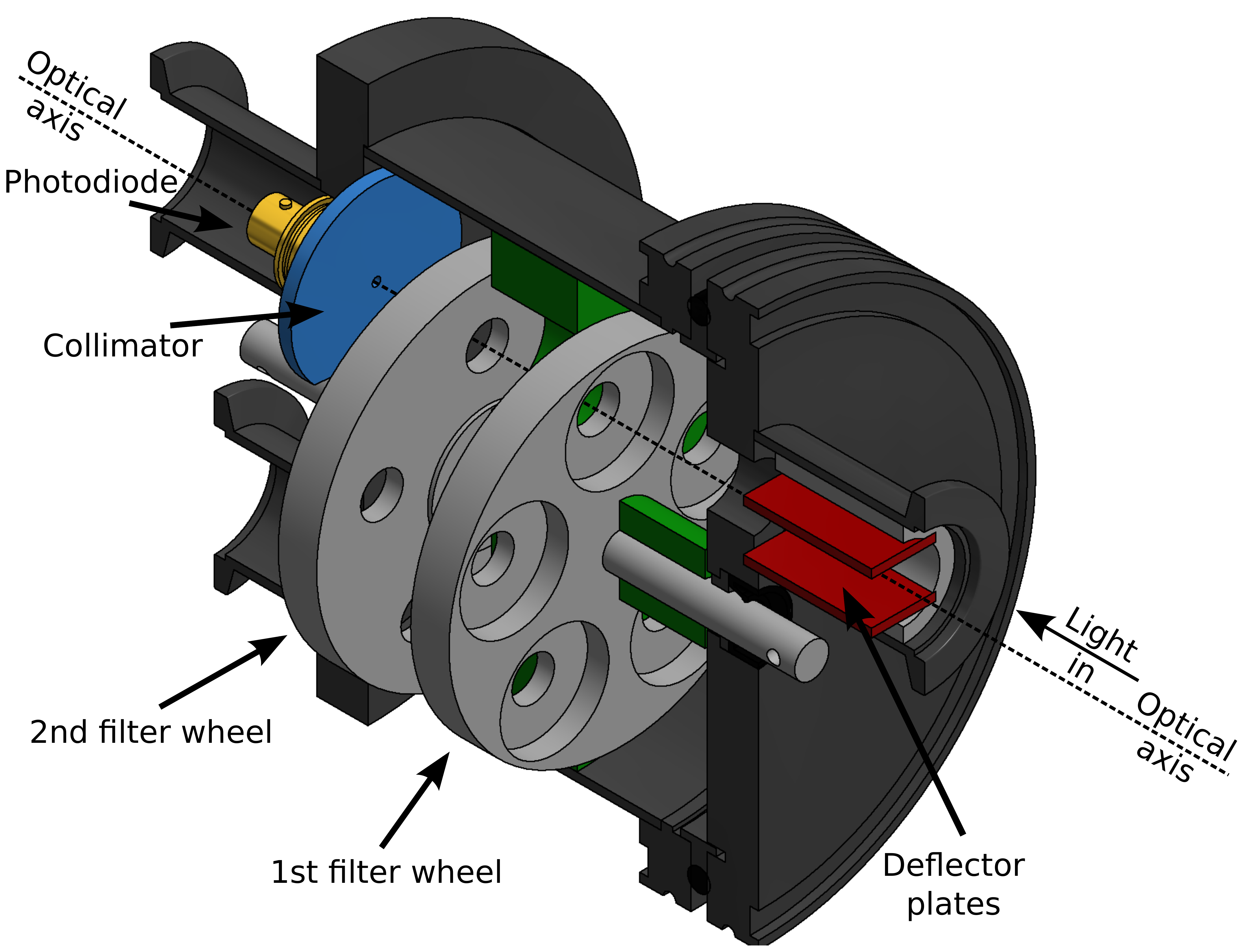}
 \caption{\label{fig:meter} Main parts of the spectral irradiance meter. The meter can be connected to a VUV-spectrometer after removing the collimator and the photodiode.}%
 \end{center}
 \end{figure}

Vacuum ultraviolet spectroscopy of plasma sources is technically challenging.  Problems arise for example from vacuum requirements for the transmission of VUV-light and sensitivity of optical elements to free radicals of residual gases induced by VUV-exposure. Spectrometers, consisting of a vacuum monochromator and a VUV-sensitive photomultiplier or photodiode, are commonly used for VUV spectroscopy. Such devices provide accurate measurements of relative intensity but their calibration for absolute irradiance is very challenging \cite{McPherson_1986_VUV_spectrometer_calibration}. In fact, the calibration procedure is sensitive to the geometry and the wavelength of the light source \cite{McPherson_1986_VUV_spectrometer_calibration}. This complicates calibration with reference VUV light sources, because their effective emission volumes ($\approx1$~mm$^3$) differ significantly from typical line of sight volume of plasma sources (\textgreater1~cm$^3$). Furthermore, the stability of reference light sources, e.g. deuterium lamps, is poor especially close to the cut off wavelength of the magnesium fluoride window (120--160~nm). 

Another method for absolute calibration is to use the branching ratio technique \cite{Klose_1989_Branching_ratio}. The method utilizes a branching pair of spectral lines emitted in transitions from a common excited state in the plasma. Calibration point in the VUV-range can be obtained, if the branching ratio of the lines is known, one of the transitions emits a photon in the VUV-range and the absolute emission intensity of the other transition can be measured e.g. in the visible light range. Utilizing this technique is challenging, for example, due to the lack of suitable branching pairs, overlapping of emission lines and geometrical effects \cite{Dong_2011_Absolute_VUV_calibration}.

\subsection{Irradiance meter}
\label{sec:meter}

In this study the plasma VUV-irradiance is measured utilizing a photodiode and set of optical filters. When the spectral response of the diode, transmittances of the filters and the effective area of the photodiode are known, the irradiance can be calculated from the measured current signal. The total VUV-emission at 80--200/350~nm wavelengths was measured by using UV-FS/BK7 window as an optical filter \cite{fusion_VUV_diagnostic,Morozov_2008_electron_beam_VUV_study}. The VUV-emission in the characteristic wavelength ranges of hydrogen was determined by using optical band pass filters. This method is robust, because the geometrical effect can be understood straightforwardly and, in principle, the response of the diode and transmittances of the filters can be determined separately. The drawback of the method is that there are more sources of uncertainty and accurate determination of the effective transmittances requires knowledge about the plasma emission spectrum due to non-uniform transmission of the filters.

The design of the VUV-irradiance meter is presented in Fig. \ref{fig:meter}. The device consists of a vacuum chamber, two rotatable wheels for optical filters, a VUV-sensitive photodiode and high voltage plates for deflecting charged particles (possibly) incident from the plasma.  The wheels are rotated externally via vacuum feedthrougs. The meter can be connected to different types of vacuum systems and spectrometers through DN25KF flanges. The photoelectric current was measured from a photodiode (IRD SXUV5) by using SRS SR570 current preamplifier. The effective area of the photodiode was restricted by a \diameter 2 mm collimator aperture, because a constant response is guaranteed only at the center of the diode.

\begin{figure}
 \begin{center}
 \includegraphics[width=0.50\textwidth]{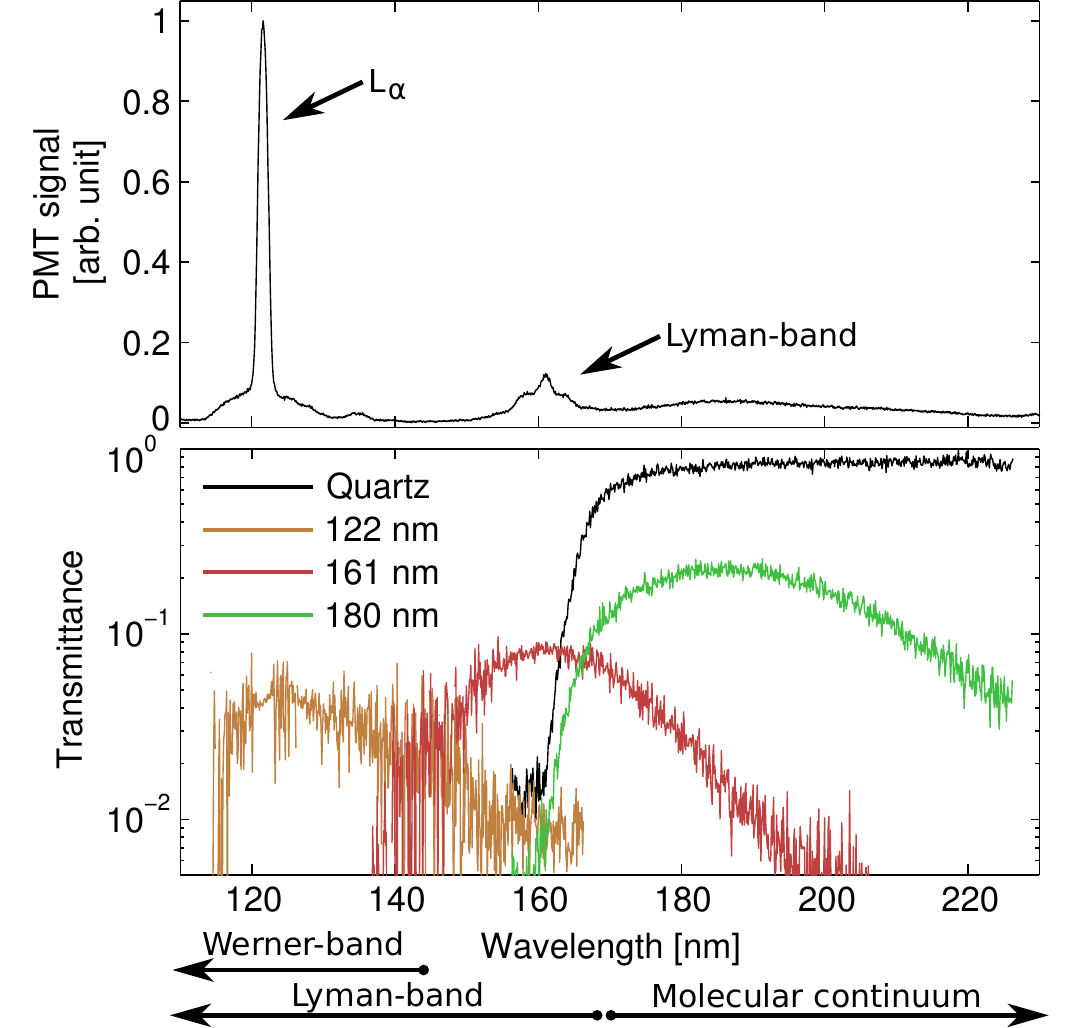}
 \caption{\label{fig:spectrum} (a) VUV-spectrum of the studied microwave discharge (b) Measured transmittances of the optical filters and the UV fused silica (UVFS) window. The spectrum is not calibrated for spectral response.}%
 \end{center}
 \end{figure}

Three different bandpass filters corresponding to the most intense emissions of the hydrogen plasma were used in the second filter wheel. These filters (122 nm / 20 nm FWHM, 161 nm / 20 nm FWHM and 180 nm / 40 nm FWHM) were used to study the emission by the lowest electronic transitions of hydrogen atom and molecule, namely Lyman-alpha (2P$\rightarrow$1S) of the hydrogen atom, Lyman-band singlet transition ($B^1\Sigma^+_u \rightarrow X^1\Sigma^+_g$) of the hydrogen molecule and molecular continuum triplet transition (a$^3\Sigma^+_g \rightarrow $b$^3\Sigma^+_u$) of the hydrogen molecule. A typical VUV-spectrum of the microwave discharge is presented in Fig. \ref{fig:spectrum} together with the measured transmittances of the filters.

\begin{figure}
 \begin{center}
 \includegraphics[width=0.50\textwidth]{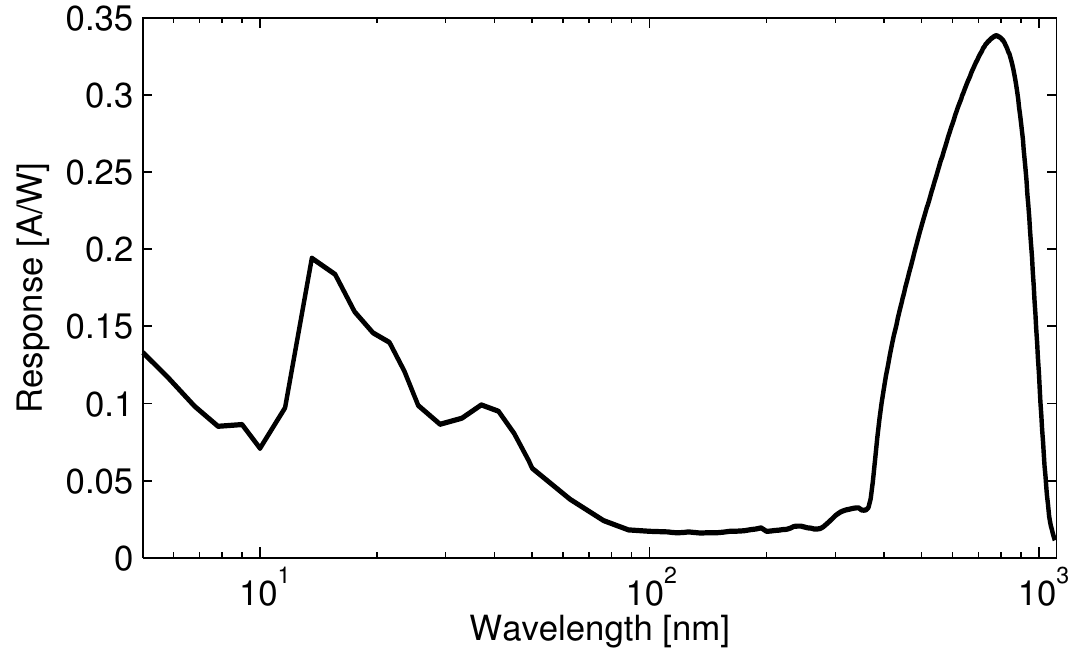}
 \caption{\label{fig:response} The spectral response of SXUV photodiode.}%
 \end{center}
 \end{figure}

The first filter wheel included a 3 mm thick uncoated UV fused silica and BK7 windows, which were used for determining the total VUV-emission. The response of the photodiode is almost constant in the range of 80-350 nm (Fig. \ref{fig:response}). On the other hand, the silica and BK7 windows are transparent in the range of 170--2000~nm and 350--2000~nm respectively. Hence, the total VUV-irradiance in the range of 80--170~nm/80--350~nm is linearly proportional to the difference of signals measured without and with the silica/BK7 windows.

The first wheel included also an aluminium plate, which was used for the measurement of background signals caused by the filters being somewhat transparent to visible light and by scattering inside the meter. Neither the visible light nor the scattering caused any measurable background signal in the case of the microwave discharge. However, thermionic electrons emitters, e.g. hot filaments can cause significant background as mentioned in Ref. \cite{komppula_NIBS_2012}. Exposure to VUV-light induces fluorescence in common optical window materials. However, the possible fluorescence does not affect the results of the measurements, because the photodiode covers only a minimal space angle of the total fluorescent emission.

All of the optical elements are sensitive to long term irradiation by VUV-light. This is especially true for the transmittance of windows and substrates of the optical filters close to their cut-off wavelengths. Therefore, it was necessary to determine the transmittances of the filters both before and after the irradiance measurements. This was done by connecting a VUV-spectrometer (McPherson Model 234/302 and ET Enterprises 9406B photomultiplier tube) to the outlet of the irradiance meter and using the plasma source described in the following section as a light source.

The average transmittances of the band pass filters were determined by comparing an integrated spectrum with and without the filter in the determined wavelength range. The wavelength ranges were determined to correspond to natural ranges of the emission bands, i.e. 115--145 nm (122~nm filter), 145--170 nm (160~nm filter) and 170--240 nm  (180~nm filter). Due to the overlap of filter bands their transmittances in adjacent ranges were also determined in order to make corresponding corrections to the measured signals. 

Analysis of different emission bands covered by the filters is necessary for plasma diagnostics purposes \cite{Komppula_2015_VUV_diagnostics}. Based on the spectral analysis it can be argued that the 115--145~nm range consists of Lyman-alpha emission together with Werner- and Lyman-bands, the 145--165~nm range consists dominantly (\textgreater80\%) of the Lyman-band emission and the 170--240~nm range purely of the molecular continuum emission. Due to vibronic excitation of the hydrogen molecule, the measured ranges do not completely cover the emission bands. The origin of the light emission in different wavelength ranges can be studied theoretically by producing synthetic spectra. The first order approximation is that the change of vibrational levels in electronic excitations follows Franck-Condon factors \cite{Fantz_2006_Franck_condon}. Such calculations depend only on the vibrational temperature of the neutral gas which is typically 100--10000~K in low temperature plasmas. Using this range of vibrational temperature it was calculated in Ref. \cite{Komppula_2015_VUV_diagnostics} that the wavelength range of 145--170~nm covers 38--43\% of the total Lyman-band emission and the range of 170--240 nm 59--62\% of the molecular continuum emission. It was also calculated in Ref. \cite{Komppula_2015_VUV_diagnostics} that 7--14\% of the measured emission in the range of 170--240 nm is caused by cascade from the upper states. Furthermore, it can be calculated that range of 115--145~nm covers 30--34\% of Lyman-band emission and 49--53\% of Werner-band emission.

Comparison of the measurements performed with UV-FS/BK7 windows and band pass filters is non-trivial. This is because the band pass filters do not cover the entire VUV-spectrum (as discussed above) and there is a significant overlapping of different emissions in 122~nm filter range. Electron impact excitation cross sections imply that the majority of VUV-emission of molecules is caused by Lyman-band ($B^1\Sigma^+_u \rightarrow X^1\Sigma^+_g$), Werner-band ($C^1\Sigma^+_u \rightarrow X^1\Sigma^+_g$) and molecular continuum (a$^3\Sigma^+_g \rightarrow $b$^3\Sigma^+_u$) transitions. The total emissions of Lyman-band and molecular continuum can be calculated from the measured signals applying the fractions explained in the previous paragraph. The electron impact excitation rate coefficient to $C^1\Sigma^+_u$ state is 0.5--0.8 times the electron impact excitation rate coefficient to $B^1\Sigma^+_u$ state, if the electron temperature is in the range of 3--25~eV. This information together with the synthetic spectrum yields that the Werner-band emission in the range of 115--145~nm is 0.57-1.0 times the measured Lyman-band emission in the range 145--170~nm. Altogether this means that the molecular emission rate in the range 115--145~nm is 2.1--2.8 times the measured rate of the Lyman-band emission in the range of 145--170~nm. The remaining measured signal in the range of 115--145~nm is caused by Lyman-alpha emission. Furthermore, it can be calculated that the ranges 115--170~nm covers 81--91\% of total excitations to $B^1\Sigma^+_u$ and $C^1\Sigma^+_u$ states.  In other words the Lyman- and Werner-band emission which is not covered by the measurements in the range of 115--170~nm (covered by 122~nm and 160~nm filters) can be estimated to be 0.3--1.0 times the measured signal in the range of  145--170~nm.

\subsection{Signal analysis}
Irradiance\footnote{This study uses photometric notation defined in Ref. \cite{CRC_91}.} $E$ at the photodiode can be calculated from the measured current $i$ with
\begin{equation}
 E=\frac{i}{A_d R\tau_{\rm f}}\ \text{[W/m}^2\text{]},
\end{equation}
where $A_d$ is the active area of the photodiode, $R$ [A/W] is the diode response and $\tau_{\rm f}$ is the transmittance of the optical filter. 

The plasma emission spatial profile is typically homogeneous and uniform in the line of the sight volume, which allows reasonably accurate determination of the average volumetric emission rate from this volume. However, it is emphasized that the line of sight volume of plasma sources can not be treated as a point source. This is because achieving detectable level of light emission requires the meter to be placed near the light source and dimensions of the line of sight volume are finite. Consequently, the geometric transmission probability of the photons to the detector varies within the line of sight volume. The most accurate method for determining the probability is to use a Monte-Carlo simulation. An analytical expression can be used in case of simple geometry, where diameters of the photodiode and light emission apertures are small in comparison to the distance between the photodiode and the plasma source. An expression for the volumetric radiant emission power of isotropic and homogeneous plasma in the line of sight volume is derived in Appendix A and can be written as
\begin{equation}
\label{eq:simple_volumetric emission}
\frac{\rmd \Phi_{\rm e}}{\rmd V}=\frac{4 \pi (D+\frac{L}{2})^2}{A_{\rm d} A_{\rm e} L(1+\frac{L}{D}+\frac{1}{3}\frac{L^2}{D^2})}\frac{i}{R\tau_{\rm f}},\ \text{[W/m}^3\text{]},
\end{equation}
where $D$ is the distance between the photodiode and the light emission aperture (referred as extraction aperture in the case of ion sources), $A_d$ is the area of the photodiode, $A_e$ is area of the extraction aperture, $L$ is the length of the plasma volume, $i$ is the measured photo diode current, $R$  is the diode response and $\tau_f$ is the transmittance of the optical filter. For the studied geometry the relative difference of the results given by Eq. \ref{eq:simple_volumetric emission} and Monte-Carlo simulation is \textless1\%. The main uncertainty of Eq. \ref{eq:simple_volumetric emission} is caused by the plasma volume, for which the detector is only partially visible.

The total volumetric radiant emission power $\Phi_{\rm tot}$ can be estimated by scaling the measured average volumetric emission power of the line of sight volume with the plasma chamber volume. The simplest approximation is to assume homogeneous and isotropic emission profile in the entire plasma chamber, which results to 
\begin{equation}
\label{eq:simple_total emission}
\Phi_{\rm tot}=\frac{\rmd\Phi_{\rm e}}{\rmd V}V=\frac{4 \pi (D+\frac{L}{2})^2}{A_d A_e L(1+\frac{L}{D}+\frac{1}{3}\frac{L^2}{D^2})}\frac{i}{R\tau_{\rm f}}A_c L,\ \text{[W]},
\end{equation}
where $A_{c}$ is the cross-sectional area of the plasma chamber. In most cases this expression gives the maximum value of the total emission power, because the plasma density decreases towards the plasma chamber walls. It can be estimated that the corresponding error between the given maximum and emission power corresponding to more realistic plasma distribution is most often less than 50\% \cite{komppula_NIBS_2012}.

\subsection{Error analysis}

The total error of irradiance consists mainly of the uncertainties caused by the photoelectric current measurement, diode response, filter transmittance and background signals. In addition, uncertainties of plasma homogeneity must be included in the case of volumetric and total emission of the plasma.

In the case presented in this paper the uncertainties of the photodiode response and transmission of the bandpass filters dominate the total uncertainty. The manufacturer (IRD) guarantees that the photodiode response is  within the limit of $\pm$10\% standard deviation from the nominal response curve (Fig. \ref{fig:response}). It is also assumed that the uncertainty of the nominal curve is negligible in comparison to the uncertainty caused by statistical variation between individual diodes. The response is almost constant in the wavelength range of 80--250 nm, i.e. the variation of the response is negligible within the specific wavelength range limited by the optical filters in comparison to the total uncertainty of the response. The photocurrent measurement includes an uncertainty of less than $\pm 1 \%$.

Non-uniform transmittance of the optical filters sets challenges for the determination of their effective (average) transmittance. Accurate determination (better than a factor of two uncertainty) of the effective transmittance of the band pass filter can be performed by integrating the spectrally weighted transmittance over the active range. This requires the relative emission spectrum of the plasma to be known in the active range. It must be noted that the spectrum should  not vary significantly during the irradiance measurements. Furthermore, the determination of stray transmittance outside the effective range of the filter increases the uncertainty. In the measurements presented here, the partial uncertainty caused by transmittances was estimated to be 8\%. The value was determined utilizing a measured spectrum and synthetic molecular emission spectra with varying vibrational temperature (100--10000~K) and electron temperature (3--25~eV).
 
The most significant partial uncertainties, namely the photocurrent, the response of the photodiode and transmittances of the filters are independent in the presented case. Therefore, the total uncertainty can be estimated with general expression of error propagation \cite{taylor} to be $\pm$13\%. 

The accuracy of the irradiance values could be enhanced significantly by using better calibrations of the diode response and filter transmittances. However, the error analysis becomes very challenging when all uncertainties are taken into account and some of them are not  independent anymore. Therefore, it can be argued that the maximum (reasonably) achievable accuracy of the presented method is $\pm$5--10\%.

The total irradiance determined by comparing the difference of the photodiode signals measured without a filter and with a silica or BK7 window includes a larger uncertainty than the spectral irradiance measurements. This is because, the photodiode response varies in the VUV-range, the transmittance of the silica window varies in the 280--1200~nm range and the photodiode response is an order of magnitude higher in the range of 280--1200~nm than in the range of 80--280~nm. The spectral response of the photodiode varies only $\pm$10\% from the average value in the wavelength range of 80--280~nm, which allows using a constant value of response in this wavelength range. In the case of hydrogen plasma, all of the emission lines and bands are emitted in this wavelength range. However, plasmas of some elements (e.g. O, He and Ne) or ions could emit strong lines below 80 nm and, therefore, the integrated diode response weighted by the spectrum must be calculated in those cases. In the case of the microwave-driven hydrogen discharge, up to 90\% of the signal measured without a filter was caused by emission in visible and infrared range. Hence, both the estimated effective transmittance of the silica window and the stability of the light source in between the measurements without a filter and with a silica window are estimated to cause a partial uncertainty on the order of $\pm$ 10--15\%. The total uncertainty of the total VUV-irradiance measurements is estimated to be $\pm$20--25\%.

\section{The microwave ion source}
\label{sec:microwave_ion_source}
\begin{figure}
 \begin{center}
 \includegraphics[width=0.90\textwidth]{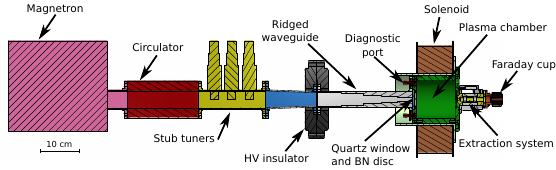}
 \caption{\label{fig:source}A schematic view of the 2.45 GHz microwave ion source. Reprinted with permission from \cite{Komppula_NIBS_2014}. Copyright 2015, AIP Publishing LLC.}%
 \end{center}
 \end{figure}

The microwave ion source used in this study is presented schematically in Fig. \ref{fig:source}. The ion source consists of a microwave coupling system, a cylindrical aluminium plasma chamber which is surrounded by a single solenoid coil, and an extraction system for positive ions. Microwaves are coupled to the plasma chamber through a rectangular waveguide system, which consists of a 2.45 GHz magnetron, circulator, three stub tuners for impedance matching and a ridged waveguide section frequently used in similar ion sources \cite{Taylor_1993_microwave_ion_source}. The vacuum of the plasma chamber is sealed from the waveguide (under atmospheric pressure) by a quartz window embedded into the back plate of plasma chamber. The quartz window is covered by a boron nitride (BN) disc. The diameter and the length of the cylindrical aluminium plasma chamber are 9.5 cm and 10 cm respectively. The ion source is typically operated with 300--1200 W of microwave power in 0.05--10 Pa\footnote{Pressure readings in this study were measured with a Pirani gauge including up to a factor 2 uncertainty \cite{TPR010}.} pressure range. The forward and reflected powers were measured from the magnetron and a power load at the circulator.

\begin{figure}
 \begin{center}
 \includegraphics[width=0.50\textwidth]{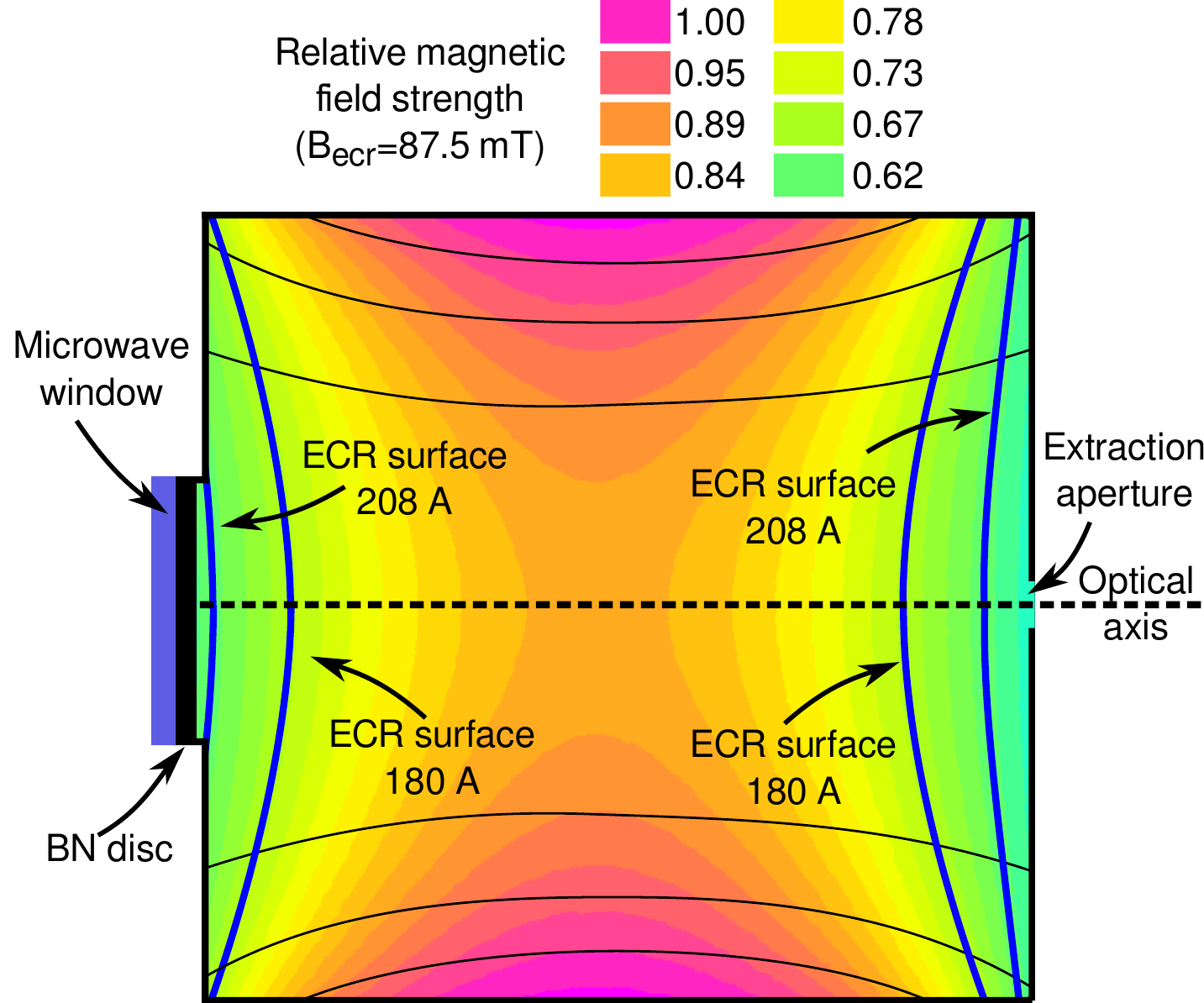}
 \caption{\label{fig:magnetic_field} Magnetic field topology and ECR surfaces corresponding coil currents of 180 A and 208 A. The magnetic field topology does not depend on coil current. The magnetic fields have been calculated with FEMM 4.2 \cite{femm}.}%
 \end{center}
 \end{figure}

Microwave discharge plasmas are typically heated with electron cyclotron resonance (ECR) heating. However, it is possible to sustain the discharge without fullfilling the ECR condition in the plasma chamber \cite{Komppula_NIBS_2014}. The ECR heating requires the electron gyrofrequency to nearly correspond with the microwave frequency, i.e. $\frac{eB}{m_{e}}\approx 2\pi f$. Hence, ECR heating with 2.45 GHz microwaves requires a magnetic field of 87.5~mT. The efficiency of the ECR heating depends on both the electric and magnetic fields. Low magnetic field gradient at the resonance surface enhances the electron energy gain by allowing the electron to spend a longer time near the optimum resonance condition. The energy gain in a single period is proportional to the electric field component perpendicular to the magnetic field. The best performance of a microwave ion source is typically achieved, when the ECR surface is directly in front of the microwave window. Figure \ref{fig:magnetic_field} represents the magnetic field of the studied ion source in terms of relative field strength as well as resonance surfaces corresponding to optimum performance (208~A coil current) and the resonance surface spanning across the whole plasma chamber (180~A).

Two different extraction apertures were used in the measurements. The ion beam, consisting mainly of H$^+$, H$_2^+$ and H$_3^+$ ions (with a small but inevitable contribution from impurities), was extracted from a 1.5 mm diameter extraction aperture to the Faraday cup located approximately at 10 cm distance from the aperture (Fig. \ref{fig:source}). The extraction system consists of a grounded puller electrode, an einzel lens and the Faraday cup. The potentials of the plasma source and the einzel lens were +10~kV and +7.5~kV respectively. A secondary electron suppression plate biased to $-150$~V was used in front of the Faraday cup.

A larger, 6 mm diameter, extraction aperture and transverse magnetic filter was used in VUV irradiance measurements. The larger aperture allowed recording a stronger VUV-signal at the irradiance meter and monochromator located 58.5 cm from the aperture. The magnetic filter was necessary to suppress an observed plasma jet pluming out of the chamber. Without the filter the plasma jet damaged optical elements and the VUV-emission of the jet caused a significant background signal affecting the measurements. The magnetic filter caused only a weak stray field in the plasma chamber.

The plasma source and the irradiance meter were separated with a gate valve and they were pumped separately with turbomolecular pumps. This allowed connecting the irradiance meter to the VUV-spectrometer without venting the plasma source. The gate valve was also used as a shutter in between the VUV-measurements.

\section{Results}

\begin{figure}
 \begin{center}
 \includegraphics[width=0.50\textwidth]{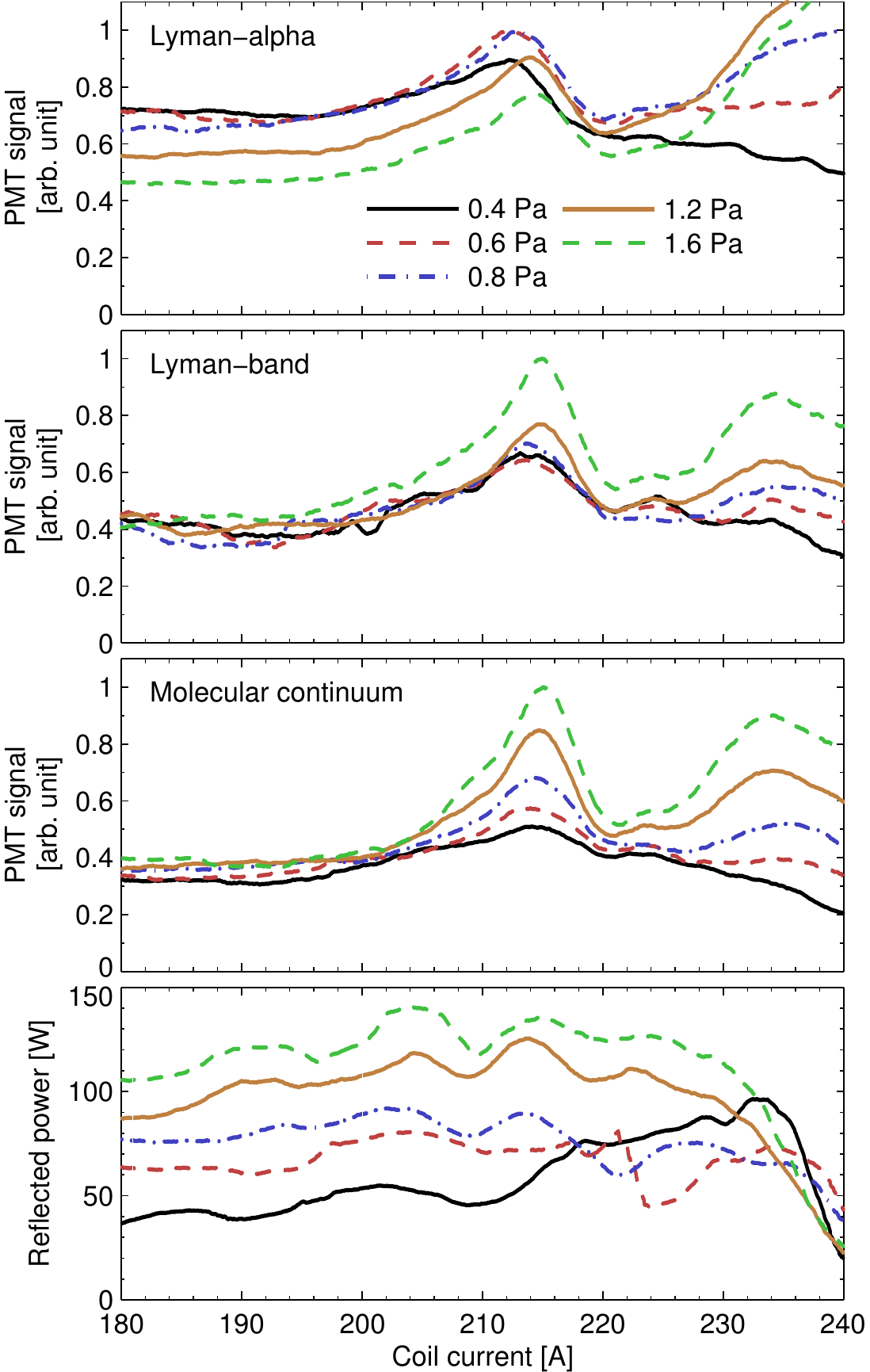}
 \caption{\label{fig:data_pressure} The VUV-emission of the hydrogen plasma within specific wavelength ranges as a function of the coil current (magnetic field strength) and neutral gas pressure at microwave power of 840 W.}%
 \end{center}
 \end{figure}

\begin{figure}
 \begin{center}
 \includegraphics[width=0.70\textwidth]{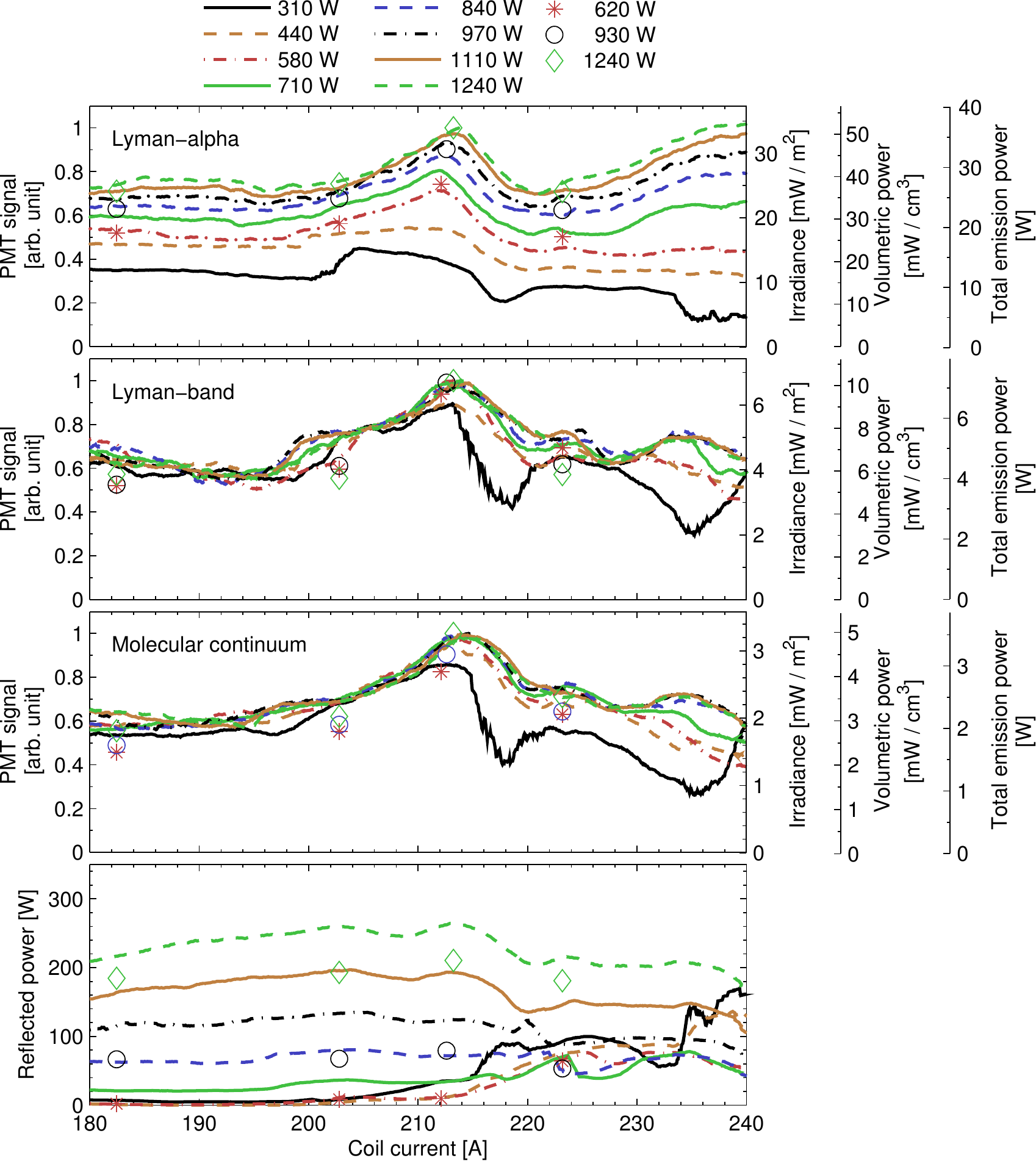}
 \caption{\label{fig:data_power} VUV-emission of the hydrogen plasma within specific wavelength ranges as a function of the coil current (magnetic field strength) and the microwave power. The neutral gas pressure in the plasma chamber was 0.6 Pa. The discrete data points are measured with the irradiance meter and are projected to the secondary vertical axes i.e. irradiance, volumetric emission power and total emission power (on the right), while the continuous lines correspond to the PMT-signals projected to the primary vertical axis (left).}%
 \end{center}
 \end{figure}

The VUV-emission was measured as a function of injected microwave power, neutral gas pressure and magnetic field strength. The most challenging aspect of the multi-dimensional parameter scan was the fact that the source performance depends on the optimization procedure. The performance is sensitive to impedance matching with the stub tuners, which partially explains the variation of the reflected microwave power as a function of the given source parameters. An unknown fraction of the microwave power was absorbed by the waveguide system. The significance of the absorption depends on the impedance matching, assembly of the waveguide system and plasma parameters. Thus, the most reliable method to perform systematic studies was to minimize the reflected power by adjusting the stub tuner positions under specific source parameters and use the corresponding tuner settings in all following measurements. The source parameters for the optimization were 600 W microwave power and 0.6~Pa neutral gas pressure at optimum magnetic field (208 A coil current). Such optimization provided good source performance for varying coil current at $<700$~W microwave power. The measurements were performed after manual cleaning of the plasma chamber followed by several hours (hydrogen) plasma cleaning to minimize the level of residual gases.

To ensure the stability of the optical filters, the absolute irradiance was measured only for a limited number of source parameter combinations around the maximum of VUV-emission. A systematic study of the VUV-emission was performed with the VUV-spectrometer, which is more stable over time. Due to the 20 nm bandwidth of the optical filter, the Lyman-alpha signal measured with the irradiance meter was affected also by the Werner-band and Lyman-band emissions, which was not included in the signal measured with the spectrometer. In addition, the beam current was measured within the same range of source parameters through a different extraction aperture as explained in Section \ref{sec:microwave_ion_source}.

The VUV-emission measured as a function of the neutral gas pressure and magnetic field strength (coil current) is presented in Fig. \ref{fig:data_pressure} and as a function of microwave power and magnetic field strength in Fig \ref{fig:data_power}. Figure \ref{fig:data_power} includes data obtained with both, the irradiance meter and spectrometer at identical source parameters. The molecular VUV-emission is almost independent of the microwave power, while the atomic emission (Lyman-alpha) is proportional to the discharge power. The dependence of the VUV-signals on the microwave power is not sensitive to the magnetic field strength. The maximum emission is observed consistently at the same coil current corresponding to the magnetic field configuration displayed in Fig. 4. The absolute irradiance of Lyman-alpha is approximately five times stronger than the Lyman-band irradiance. On the other hand, the irradiance of the Lyman-band is twice as strong as the irradiance of the molecular continuum emission. Varying the neutral gas pressure affects each emission band differently. There is an optimum pressure range (0.6--0.8~Pa) for Lyman-alpha emission. This pressure window is independent of the the magnetic field strength. The Lyman-band emission is almost independent of pressure below 1.2 Pa while the molecular continuum emission is proportional to the pressure only at the optimum magnetic field.

\begin{figure}
 \begin{center}
 \includegraphics[width=0.50\textwidth]{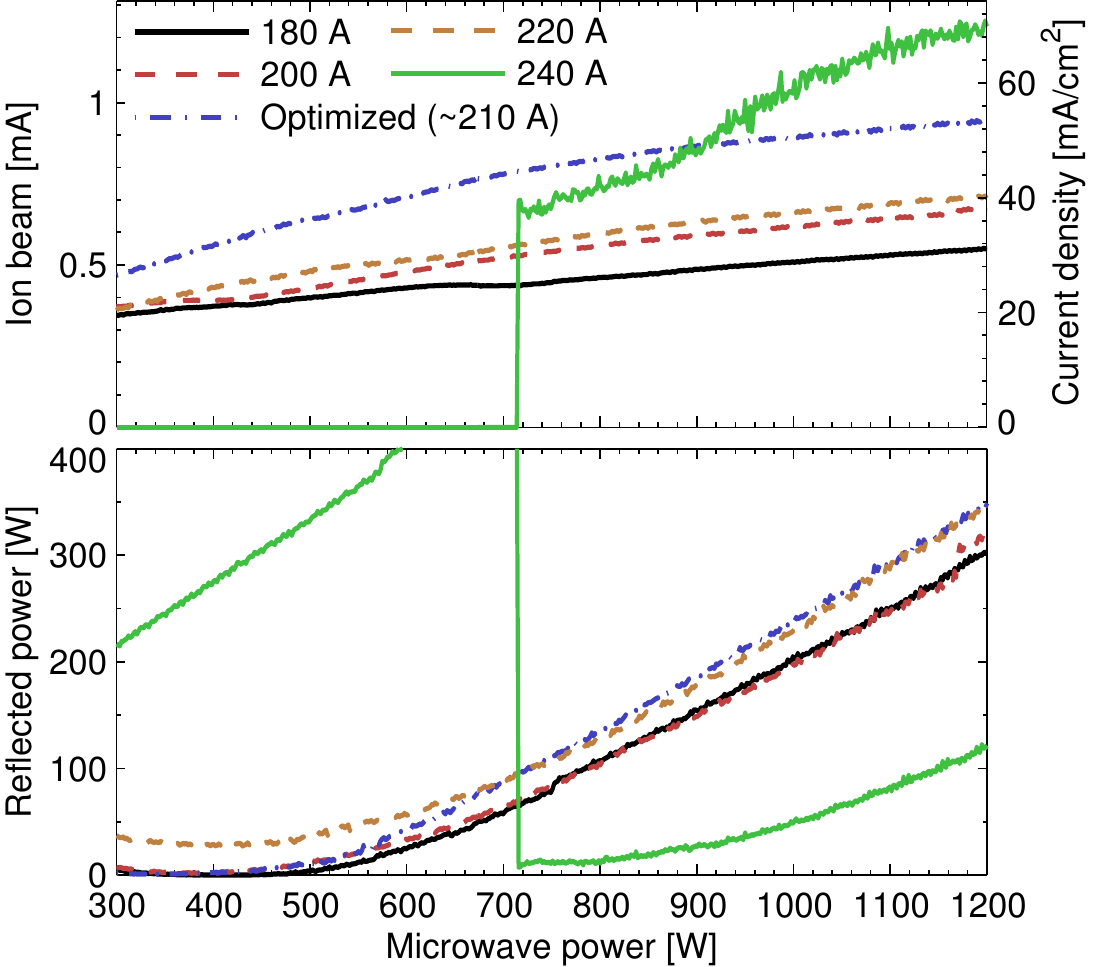}
 \caption{\label{fig:beam_power} The extracted ion beam currents as a function of the microwave power and coil current (magnetic field strength). The neutral gas pressure in the plasma chamber was 0.7~Pa.}%
 \end{center}
 \end{figure}

\begin{figure}
 \begin{center}
 \includegraphics[width=0.50\textwidth]{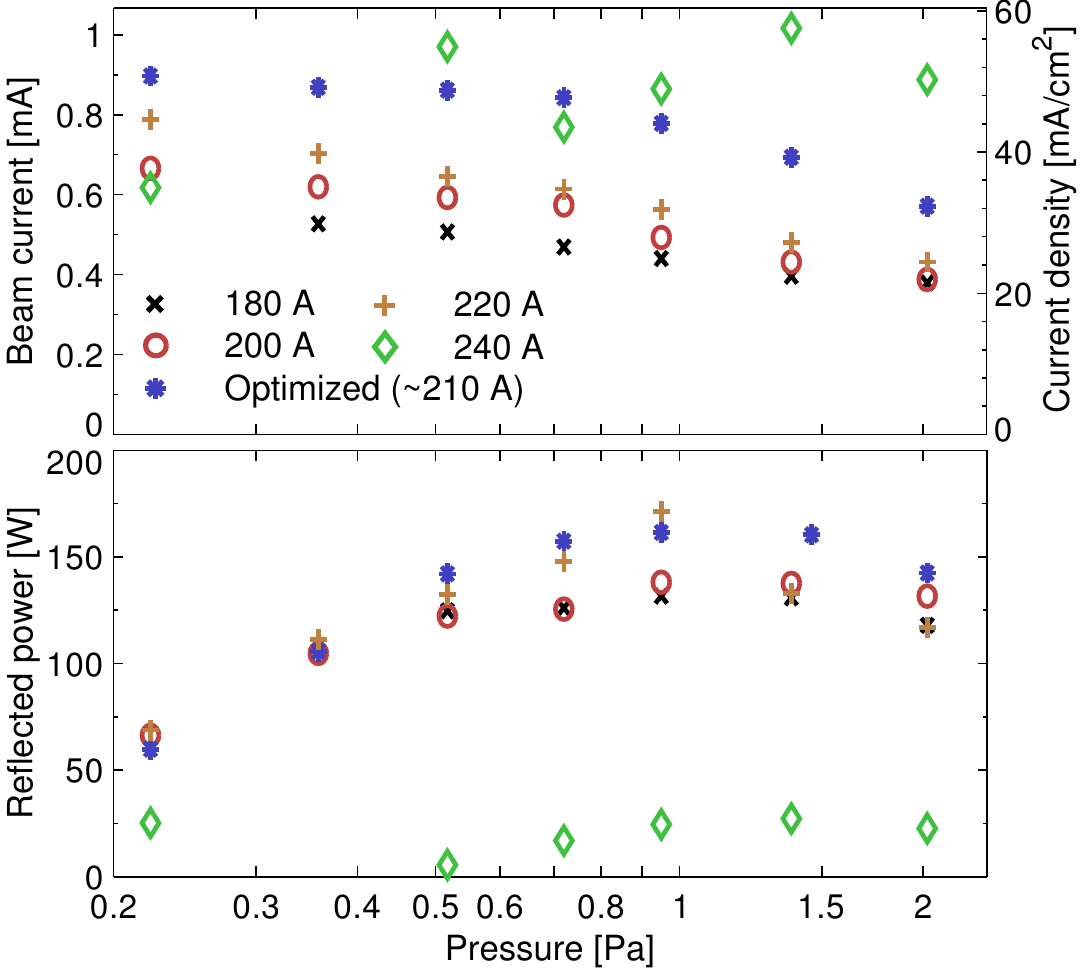}
 \caption{\label{fig:beam_pressure} The extracted ion beam currents as a function of the neutral gas pressure and coil current (magnetic field strength) at microwave power of 840 W.}%
 \end{center}
 \end{figure}

\begin{figure}
 \begin{center}
 \includegraphics[width=0.50\textwidth]{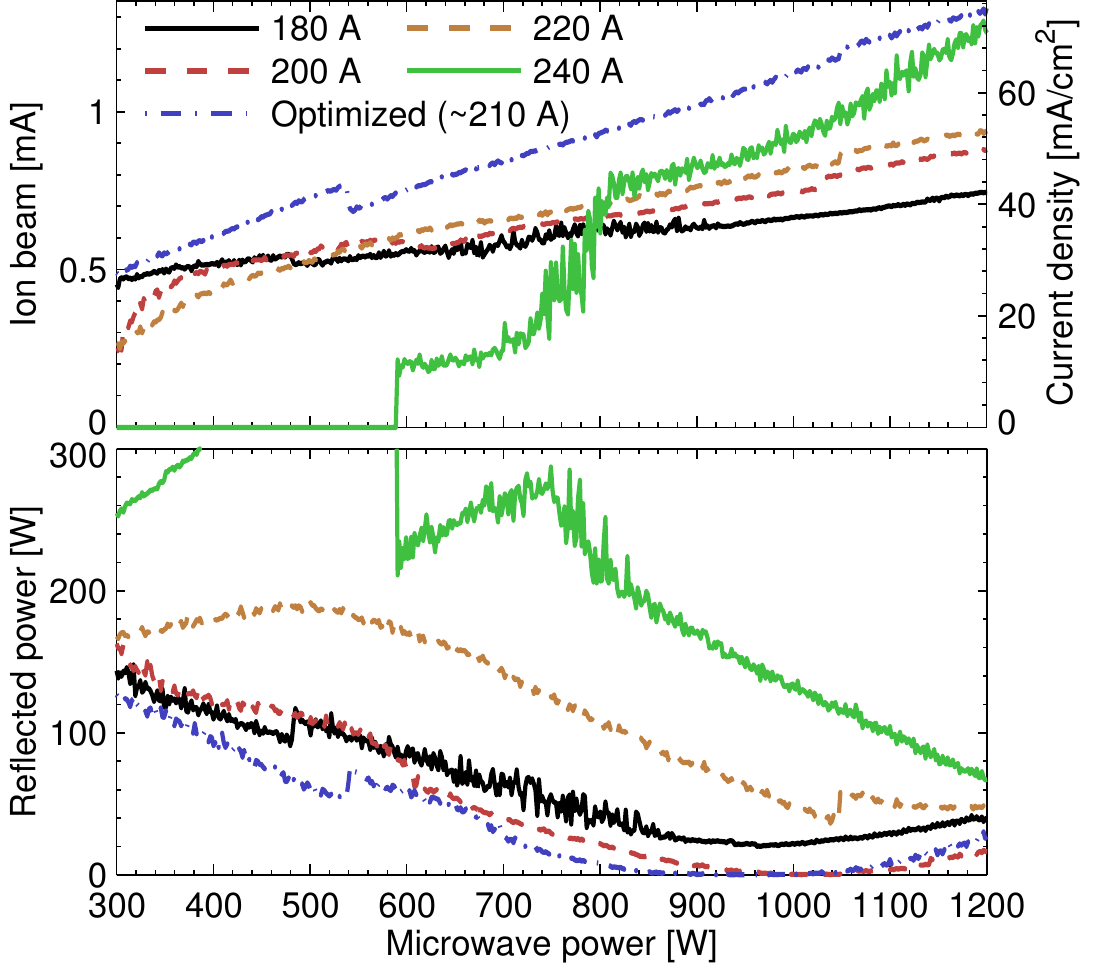}
 \caption{\label{fig:beam_high_power} The extracted ion beam current as a function of the microwave power and coil current (magnetic field strength). The reflected power was minimized at 1000 W microwave power and 210 A coil current. The neutral gas pressure in the plasma chamber was 0.5 Pa.}%
 \end{center}
 \end{figure}

Figures \ref{fig:beam_power}, \ref{fig:beam_pressure} and  \ref{fig:beam_high_power} present the extracted ion beam current as a function of various source parameters. Similar to the Lyman-alpha emission, the extracted ion beam current is proportional to the microwave power. In both cases 30\% more signal is achieved at the optimum magnetic field in comparison to lower magnetic field strengths. The most significant difference in the behaviour of the Lyman-alpha signal and extracted beam current is observed as a function of the neutral gas pressure. The optimum pressure for Lyman-alpha emission is 1.2--1.6 Pa, while the ion beam current is inversely proportional to the pressure within the measured range. The saturation of the ion beam current and Lyman-alpha emission at high microwave power is caused by the impedance tuning. This is confirmed by  the ion beam current data measured with optimizing the impedance matching at discharge power of 1000 W (Fig. \ref{fig:beam_high_power}) resulting to higher beam current e.g. at maximum power.

\begin{table}
\caption{\label{tab:total_emission_power}Total VUV-emission powers estimated from the difference of the signals without a filter and through BK7 (a) or UV fused silica (b) windows and (c) from sum of the emissions presented in Fig. \ref{fig:data_power}. Due to technical reasons the data in (a) and (b) were measured at different times than the data in part (c). The values include less than 50\% systematic uncertainty caused by the assumption of homogeneous plasma (see \cite{komppula_NIBS_2012} for further discussion).} 
\begin{indented}
\item[]\begin{tabular}{@{}lrrrl}
\br
\multicolumn{5}{c}{Discharge power}\\
& 0.6 & 0.9 & 1.2 & kW\\
\mr
\multicolumn{1}{r}{\textbf{(a)}} & \multicolumn{4}{l}{\textbf{Signal difference}}\\
 & \multicolumn{4}{l}{\textbf{(BK7, 80--350~nm)}}\\
180 A 	&   21 & 27 & 32 & W\\
200 A 	&   36 & 40 & 42 & W\\
Optimum &   46 & 52 & 57 & W\\
220 A 	&   34 & 41 & 44 & W\\
\multicolumn{1}{r}{\textbf{(b)}} & \multicolumn{4}{l}{\textbf{Signal difference}}\\
 & \multicolumn{4}{l}{\textbf{(UVFS, 80--170~nm)}}\\
180 A 	&   18 &  23 &  28 & W\\
200 A 	&   30 &  34 &  35 & W\\
Optimum &   40 &  45 &  49 & W\\
220 A 	&   30 &  35 &  38 & W\\
\multicolumn{1}{r}{\textbf{(c)}} & \multicolumn{4}{l}{\textbf{Sum of emission bands}}\\
 & \multicolumn{4}{l}{\textbf{(115--240~nm)}}\\
180 A 	&   24 &  28 &  32 & W\\
200 A 	&   27 &  31 &  33 & W\\
Optimum &   37 &  43 &  47 & W\\
220 A 	&   26 &  29 &  32 & W\\

\br
\end{tabular}
\end{indented}
\end{table}

The average volumetric emission power  in the line of sight plasma volume and the total emission power in the given wavelength ranges were calculated from the absolute irradiance measurements by using Eqs. \ref{eq:simple_volumetric emission} and \ref{eq:simple_total emission}. The results are shown in Fig. \ref{fig:data_power} as described above.  The total VUV-emission can be estimated in two different ways: directly from the difference of the signals recorded without a filter and through the BK7 or UV fused silica window, (Table \ref{tab:total_emission_power}a and \ref{tab:total_emission_power}b) and from the sum of the measured emission bands (Table \ref{tab:total_emission_power}c) combined with the estimated emission outside the transmission bands of the optical filters. Correction factors for such undetected molecular emission are discussed in Section 2. Thus, it can be argued that the value of total VUV-emission measured with silica and BK7 windows should be (theoretically) at least 2--7~W greater than the sum of the measured emission bands. These estimates assume that other Lyman-series transitions except the Lyman alpha are not taken into account. However, the results obtained through different methods (Table \ref{tab:total_emission_power}) are barely inside uncertainty limits discussed in Section 2.2 probably due to some unknown systematic variable or limited reproducibility (especially outside the optimum magnetic field).

\begin{figure}
 \begin{center}
 \includegraphics[width=0.50\textwidth]{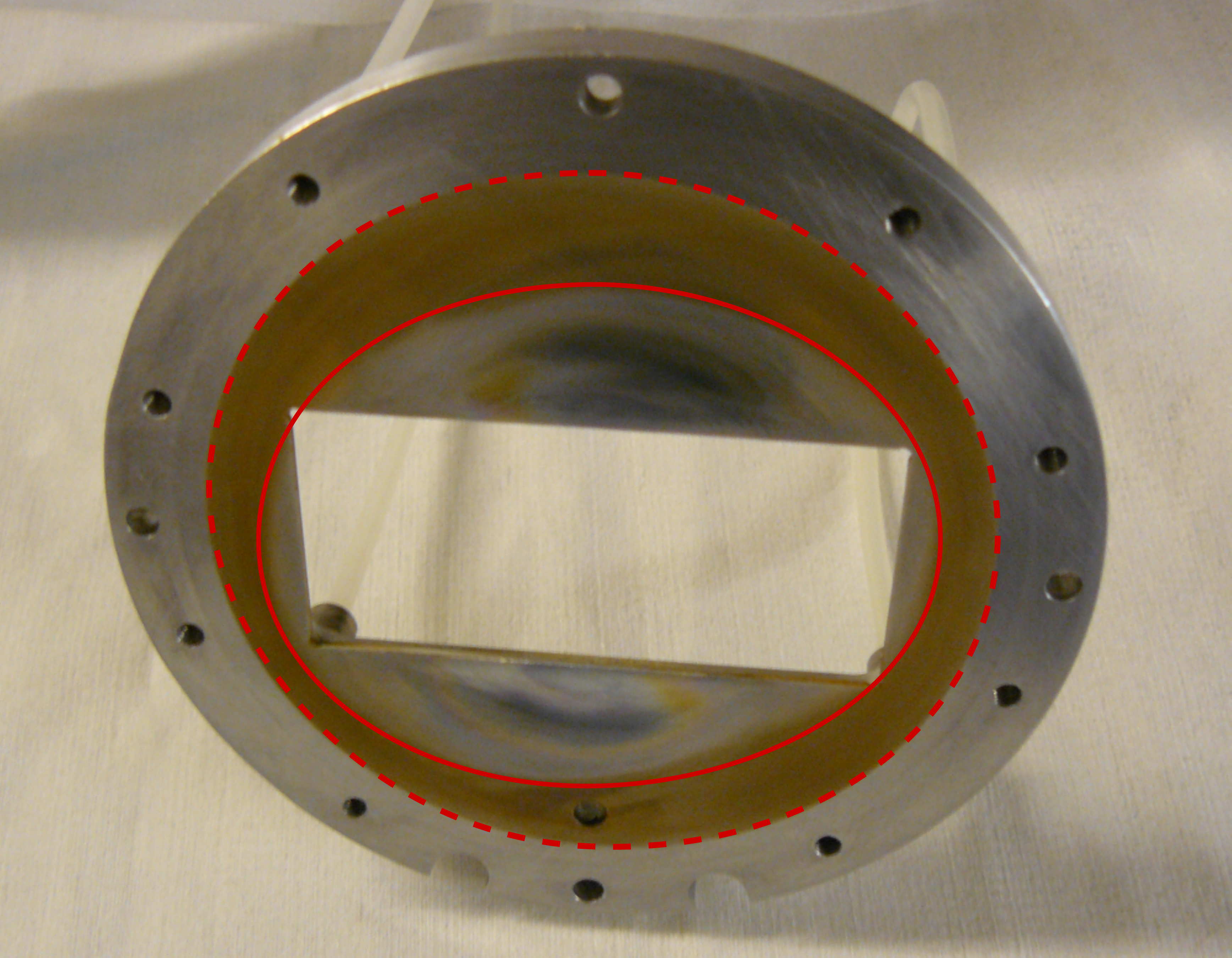}
 \caption{\label{fig:backplate} The back plate of the plasma chamber after sustaining the hydrogen discharge for several tens of hours. The area subjected to plasma bombardment (solid line) and the cross sectional area of the plasma chamber (dashed line) are highlighted in the figure.}%
 \end{center}
 \end{figure}

The total emission power values given in Fig. \ref{fig:data_power} and Table \ref{tab:total_emission_power} are maximum values based on the assumption that the plasma is homogeneous and uniformly distributed in the plasma chamber. It can be argued that the clean area on the back plate of the plasma chamber shown in Fig. \ref{fig:backplate} represents the cross sectional area of high density plasma distribution. This area covers approximately 70\% of the total cross sectional area of the plasma chamber. Therefore, it can be estimated that real total emission power values differ less than $50$\% from the values given in Fig. \ref{fig:data_power} and Table \ref{tab:total_emission_power}.

\begin{table}
\caption{\label{tab:emission_numbers}Photon emission rates of different transitions. The values are calculated from the measured irradiance values of  Fig. \ref{fig:data_power} at 930 W power and optimum resonance magnetic field. The variation of the values correspond to variation of correction coefficients discussed in section \ref{sec:meter}. The variation of total values includes 50\% uncertainty caused by the unknown plasma emission profile.}
\begin{indented}
\item[]\begin{tabular}{l cc}
\br
&Volumetric & Total\\
&[1/cm$^3$s] & [1/s] \\
\mr
Ly$_\alpha$ & \multirow{2}{*}{1.0--1.5$\times10^{16}$} & \multirow{2}{*}{3.2--9.3$\times10^{18}$}\\
($2P\rightarrow 1S$) & & \\
Lyman-band & \multirow{2}{*}{1.3--1.6$\times10^{16}$} & \multirow{2}{*}{4.2--10$\times10^{18}$}\\
($B^1\Sigma^+_u \rightarrow X^1\Sigma^+_g$) & & \\
Werner-band & \multirow{2}{*}{0.7--1.0$\times10^{16}$} & \multirow{2}{*}{1.0--8.1$\times10^{18}$}\\
($C^1\Sigma^+_u \rightarrow X^1\Sigma^+_g$) & & \\
Molecular continuum & \multirow{2}{*}{1.3--1.4$\times10^{16}$} & \multirow{2}{*}{4.2--8.8$\times10^{18}$}\\
($a^3\Sigma^+_g \rightarrow b^3\Sigma^+_u$) & & \\
\br
\end{tabular}
\end{indented}
\end{table}

Total emission of different transitions can be calculated from measured emissions in different ranges by using correction coefficients as explained in section \ref{sec:meter}. This is done in table \ref{tab:emission_numbers} where volumetric and total emissions of different transitions are presented. The calculation implies that the total volumetric photon emission rate in the VUV-range is on the order of $10^{16}$--$10^{17}$~1/cm$^3$s. The total photon emission rate integrated over the whole plasma chamber volume is on the order of $10^{19}$~1/s.  The average number of excitations / photon emissions for each injected molecule can be estimated by comparing the total photon emission rate of specific emission bands to the gas feed rate\footnote{The measured gas feed rate of 1.2~sccm corresponds to $5.3\times10^{17}$ hydrogen molecules per second.} This yields that each injected hydrogen molecule emit 2--20 photons in each emission band.

The molecular photon emission can be used as a diagnostics for plasma processes by comparing the rate coefficients of different plasma processes to measured emission rates as explained in Ref. \cite{Komppula_2015_VUV_diagnostics}. This is because (1) the ratio of specific rate coefficients does not depend on the electron or neutral densities, (2) the ratio is only weakly sensitive to the electron temperature, (3) the measured Lyman-band and molecular continuum emissions are not overlapping with other transitions and (4) the origin of the molecular excitations can be understood straightforwardly. The Lyman-band emission is proportional to molecule ionization, production of high vibrational levels $X^1\Sigma^+_g(v$\textgreater5$)$ and molecule dissociation via vibrational continuum. Furthermore, the molecular continuum emission is proportional to molecule dissociation via lowest triplet state ($b^3\Sigma_u^+$) and production of metastable states ($c^3\Pi_u$). Ratios of the rate coefficients (from Ref. \cite{Komppula_2015_VUV_diagnostics}) and calculated volumetric rates for different molecular plasma processes are presented in table \ref{tab:ratios} by assuming  that the electron temperature and molecule vibrational temperature are in the typical range of low temperature plasmas (3--15~eV and 100--10000~K). The results imply that each molecule is both ionized and dissociated several times during its passage through the plasma chamber.

\begin{table}
\caption{\label{tab:ratios} Ratios of rate coefficients (plasma process vs. Lyman-band or molecular continuum emission) and volumetric reaction rates in the line of sight volume. The ratios of rate coefficients for molecular processes are from Ref. \cite{Komppula_2015_VUV_diagnostics}  and their variation to the ranges of unknown plasma parameters ($T_e=$3--15~eV and $T_{vib}=$100--10000~K). Volumetric emissions rates of transitions are from Table \ref{tab:emission_numbers}.}
\begin{indented}
\item[]\begin{tabular}{lcr}
& Ratio of & Volumetric rate\\
& rate coefficients & [1/cm$^3$s] \\
\br
\textbf{Lyman-band} &  & \\
Molecule ionization & 0.1--1.3 & 0.14--2.1$\times10^{16}$\\
Excitation ($B^1\Sigma^+_u$, $C^1\Pi_u$) & 1.4--1.8 & 1.8--2.8$\times10^{16}$\\
Excitation ($\nu\geq5$) ($B^1\Sigma^+_u$, $C^1\Pi_u$) &  0.8--1.1 & 1.1--1.7$\times10^{16}$\\
Dissociation ($B^1\Sigma^+_u$, $C^1\Pi_u$) & 0.2--0.3 & 0.3--0.4$\times10^{16}$\\
&\\
\textbf{Molecular continuum}&  & \\
Dissociation (via $b^3\Sigma^+_u$) & 2.2--8.5 & 2.8--12$\times10^{16}$\\
Excitation $c^3\Pi_u$ state, max & 0.6--1.8 & 0.8--2.5$\times10^{16}$\\
\br
\end{tabular}
\end{indented}
\end{table}

The maximum density of metastable molecules ($c^3\Pi_u$) was estimated from the volumetric excitation rate in Ref. \cite{Komppula_2015_VUV_diagnostics} by assuming that the plasma distribution is homogeneous across the plasma chamber, $c^3\Pi_u$ states decay without emission in molecular continuum and the maximum lifetime of $c^3\Pi_u$ states is determined by diffusion to the wall of the plasma chamber. Using similar assumptions, the production rate of 1.8$\times10^{16}$~1/cm$^3$s of the metastable molecules in the microwave discharge (Table \ref{tab:ratios}) corresponds to 6.4$\times10^{11}$~1/cm$^3$ maximum density of $c^3\Pi_u$ states . This is approximately 0.4\% of the neutral gas density. The value estimated in Ref. \cite{Komppula_2015_VUV_diagnostics} is at least an order of magnitude higher than earlier experimental value from the similar type of plasma source. This probably holds also for value presented for the microwave discharge because of large uncertainties of the assumptions.

VUV-diagnostics of plasma processes based on atomic Lyman-alpha emission is more challenging than utilizing molecular emission bands. This is because the opacity of the plasma has a significant effect on the transmission of the emitted photons through the plasma and because a significant fraction of atomic emission can originate from dissociative excitation of hydrogen molecules. Furthermore, the dominant electron impact processes with same order of magnitude cross sections are excitations to 2P state, resulting directly to Lyman-alpha emission, and excitations to metastable 2S state with a lifetime on the order of 100 ms. The dynamics of the 2S excited atoms is complex. The 2S state can decay via Lyman-alpha emission with collisional quenching by neutral particles \cite{2S_Collisional_Quenching} or via interaction with the wall of the plasma chamber. On the other hand, 2S excited atoms can be readily re-excited or ionized because the threshold energies of those processes are low and corresponding cross sections are large. 

A rough estimate for the production rate and density of 2S excited metastable atoms can be presented similar to the molecular metastable density presented in Ref. \cite{Komppula_2015_VUV_diagnostics}. This requires assuming that (1) the lowest excited states (2P and 2S) are mainly populated directly via electron impact excitation from the ground state, (2) the electron impact excitation rate from ground state to 2S is 0.2--0.6 times the electron impact excitation rate to 2P state (calculated with cross section data from Ref. \cite{Janev2003} and $T_e=$3--15~eV), (3) molecular emission in the range of 115-145~nm is 1.65--1.94 times the measured (molecular) emission in the range 145--165~nm and (4) the plasma is homogeneous and uniformly distributed into the chamber, i.e. the measured average volumetric emission is not disturbed by the opacity. Assuming that none of the 2S states decay via Lyman-alpha emission corresponds to the maximum excitation rate to 2S state, i.e. 0.6 times the measured Lyman-alpha photon emission rate. The maximum realistic lifetime of 2S states is 18~$\mu$s, which is the average diffusion time of thermal atoms to the wall of the plasma chamber\footnote{Average thermal velocity of hydrogen atoms in 300K temperature is 2700~m/s. Hence the distance of 5~cm corresponds to the time of flight of 18~$\mu$s.}. This together with the estimated maximum of total 2S excitation rate gives the maximum density of 2S atoms of $2.7\cdot10^{11}$~1/cm$^3$, which corresponds to approximately 0.2\% of the neutral gas density.

\section{Discussion}
The performances and characteristics of plasma sources, especially microwave and radio frequency driven discharges, are extremely sensitive to their designs, which complicates generalizing the obtained results. Furthermore, the performance of the studied plasma source, especially as a function of the coil current outside the optimum magnetic field, was observed to be somewhat sensitive to different assemblies of the waveguide system and level of impurities in the plasma chamber. Therefore, the accuracy of the results is not only determined by the accuracy of the measurement technique but also by the level of reproducibility. Hence, the following discussion and conclusions are limited to the absolute values of VUV-emission at the order of magnitude level, which is the novelty of the presented research, and conclusions which can be tested and reproduced in other plasma sources, e.g. by repeating the described measurements or studying the effects of mechanical changes as discussed below.

The significance of different molecular plasma processes can be estimated by comparing the number of photons emitted the corresponding transitions \cite{Komppula_2015_VUV_diagnostics}. The comparison of Lyman-band and molecular continuum photons yields information on the dissociation and ionization of hydrogen molecules. In the microwave discharge discussed here the ratio of Lyman-band emission to molecular continuum emission is 2, while in a filament drive arc discharge studied in Ref. \cite{komppula_NIBS_2012} the given ratio is 4. This implies that in the microwave discharge the fraction of inelastic collisions leading to dissociation of hydrogen molecules in comparison to molecular ionization is higher. The observed difference between the plasma sources can be explained by different electron energy distribution functions (EEDF), and consequent reaction rates reflecting the different functional shapes of electron impact cross sections\cite{Komppula_2015_VUV_diagnostics,Janev2003}. The EEDF of the microwave discharge has a higher fraction of electrons in the energy range of 7--15 eV in comparison to energy range of $E_e>15$ eV. This also means that the EEDF of the microwave discharge is less efficient for ionization, i.e. electron impact excitations dominate the energy dissipation in comparison to ionization.  

The behaviour of extracted beam and VUV-light emission as a function of the magnetic field strength gives information about the plasma-microwave interaction. The maxima of the extracted ion beam current and VUV-emission are achieved at the same magnetic field, which corresponds to the situation where the resonance surface is located in very close proximity to the microwave window. This implies that the extracted beam current (plasma density near the extraction aperture) is strongly coupled with the ionization rate (which is proportional to excitation rate \cite{komppula_NIBS_2012}) in the plasma. The magnetic field gradient is weak in the middle of the plasma chamber, which should increase the energy gain of the electrons in ECR heating. However, narrow peak of optimum magnetic field implies that efficient ECR plasma heating can occur only very close the microwave window. Such behaviour is typically observed in the microwave discharges, in which the highest plasma density is achieved close to the microwave window \cite{Ana_2014_Coupler_design}. This is probably caused by significant weakening of the microwave electric field towards the center of the plasma chamber. Such weakening is due to high plasma density damping the electric field and/or the electric field pattern being modified by the ridged waveguide \cite{Taylor_1993_microwave_ion_source}.

Absolute values of molecular continuum emission have been measured earlier from microwave discharges by Fantz et al \cite{Fantz_2000_Molecular_continuum}. They observed volumetric emission rates on the order of $10^{15}$--$10^{16}$~1/cm$^3$s, which is consistent with this study.

The maximum total density of metastable particles (2S state of atom and $c^3\Pi_u$ state of molecule) was estimated to be approximately 0.5\% of the neutral gas density. This does not exclude the effect of metastable states to plasma dynamics, because the electron impact ionization cross sections from the metastable state are at least an order of magnitude larger than the ionization cross sections from the ground states. Furthermore, the significantly lower ionization potential allows a larger fraction of electrons to ionize neutrals excited to the metastable states. However, this statement does not allow clear conclusions about the role of metastable neutrals due to larger uncertainties of the estimations and sensitivity of the processes to the electron temperature.

\section{Conclusion}
A robust and straightforward method to measure VUV-irradiance and to estimate the volumetric emission power of a plasma is presented. The hydrogen plasma of a 2.45 GHz microwave discharge was used as an example although the method can be applied also for diagnostics of other types of discharges and gases. The only limitation for the spectral measurements is the availability of bandpass filters. Most of the optical bandpass filters consist of a of metal-dielectric-metal thin-films on top of the substrate material that must be transparent to VUV-light. Substrate materials with shortest cut-off wavelengths are magnesium fluoride (MgF$_2$, 115 nm) and calcium fluoride (CaF$_2$, 120 nm). However, the total irradiance at shorter wavelengths can be determined by comparing the difference of the photodiode signals measured without a filter and with a quartz window, for example.

Approximately 8\% of the microwave power was observed to dissipate via VUV light emission in optimum (resonance) conditions and 5\% in off resonance conditions at low injected microwave power ($\leq$ 600 W). At high power the corresponding fractions were somewhat smaller. Such percentage values are less than half of those measured with a filament driven arc discharge \cite{komppula_NIBS_2012}. The predominant factors explaining the difference are probably weaker plasma confinement of the microwave discharge and microwave power dissipation in the waveguide. The confinement of primary electrons in the multi-cusp magnetic field of the filament arc discharge is good \cite{Lieberman_2005_book,Leung_1975_multicusp_confinement,komppula_NIBS_2012}, which allows the primary electrons to dissipate their energy in multiple inelastic collisions \cite{komppula_NIBS_2012}, which in turn enhances the VUV-emission of the plasma. The temperature of the ridged waveguide of the microwave discharge was observed to be 80--100~$^{\circ}$C at 1~kW power, which indicates that a significant fraction of the microwave power could be absorbed by the waveguide.

The volumetric reaction rates imply that each injected molecule experiences several reactions including ionization, dissociation and excitation to high vibrational levels via $B^1\Sigma^+_u$ and $C^1\Pi_u$ states. This underlines the role of plasma chamber surfaces to plasma dynamics, because the equilibrium is dominantly determined by reverse processes on the surfaces (e.g. Ref. \cite{Bacal_2015_negative_hydrogen_review} and references therein).

\section*{Acknowledgements}
This work has been supported by the EU 7th framework programme 'Integrating Activities -- Transnational Access', project number: 262010
(ENSAR) and by the Academy of Finland under the Finnish Centre of Excellence Programme 2012--2017 (Nuclear and Accelerator Based Physics Research at JYFL)

\appendix
\section{Volumetric light emission of the plasma}
Assuming that the light emission in the line of sight volume is isotropic and homogeneous, corresponds to a constant volumetric radiant intensity of 
\begin{equation}
\label{eq:VolumetricRadiantIntensity}
\frac{\rmd I}{\rmd V}=\frac{\rmd^2\Phi_e}{\rmd V \rmd\Omega},
\end{equation}
where $\Phi_e$ is the radiant emission power, $V$ is the (plasma) volume and $\Omega$ is the space angle. 
When the diameters of the detector and the extraction aperture are small in comparison to the distance between the detector and light emission point, the volumetric radiant intensity can be presented in the form of finite $\Delta$
\begin{equation}
\label{eq:VolumetricRadiantIntensity_finite}
\frac{\rmd I}{\rmd V}\approx\frac{\Delta\Phi_r}{\Delta V\Delta\Omega},
\end{equation}
where $\Delta\Phi_r$ is the radiant power received at the detector, $\Delta V$ is the line of sight plasma volume and $\Delta\Omega$ is the observable space angle. Assuming, that detector is completely visible from the emission point, it covers a space angle,
\begin{equation}
\label{eq:finite_space_angle}
\Delta\Omega\approx4\pi\frac{A_d}{4\pi X^2},
\end{equation}
where $A_d$ is the area of the detector and $X$ is the distance between the detector and the emission point. The simplest approximation for  the line of sight volume is cylinder, i.e. the product of the area of the extraction aperture $A_e$ and the length of the plasma chamber $L$:
\begin{equation}
\label{eq:LineOfSightVolumeSimplest}
\Delta V= A_eL.
\end{equation}
However, such approximation causes an uncertainty of approximately 20\% in the presented case. This is because in reality the line of sight volume is a truncated cone instead of cylinder. The volume of a truncated cone seen by the centre point of  the detector can be expressed as, 
\begin{equation}
\label{eq:LineOfSightVolume}
\Delta V= A_e L\left(1+\frac{L}{D}+\frac{1}{3}\frac{L^2}{D^2}\right),
\end{equation}
where $D$ is the distance between the detector and the extraction aperture.

\begin{figure}
 \begin{center}
 \includegraphics[width=0.70\textwidth]{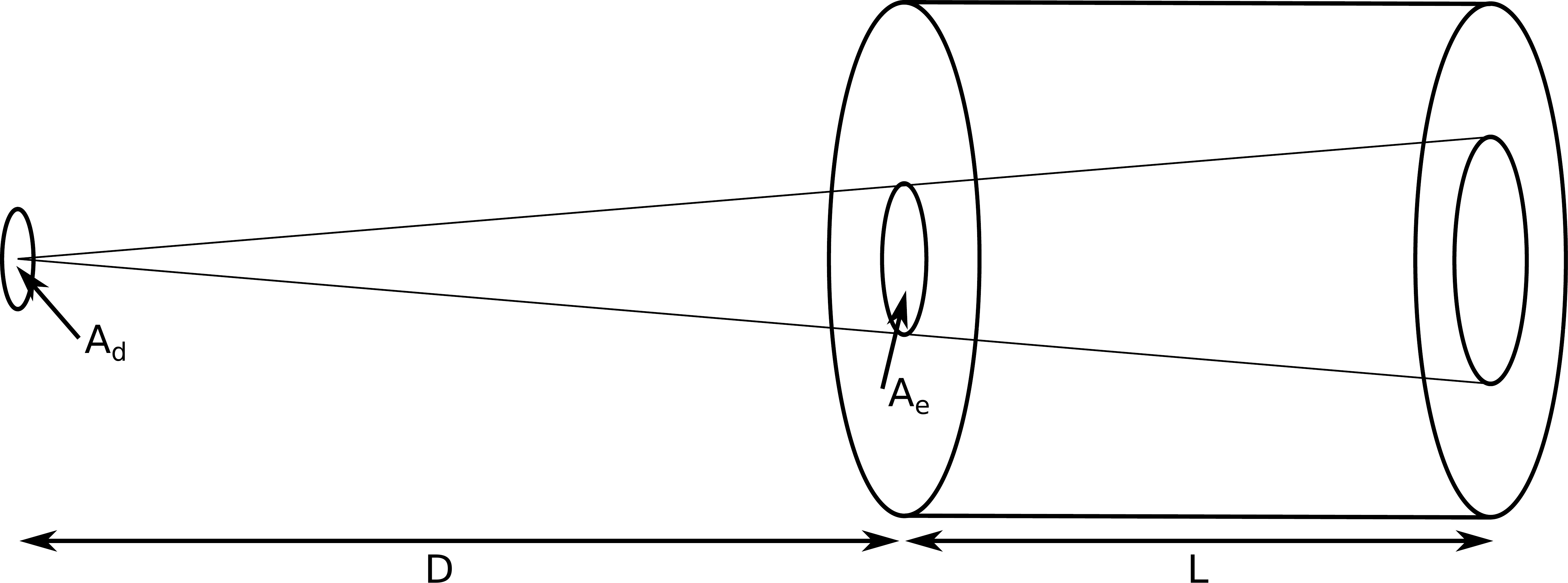}
 \caption{\label{fig:schematic} Schematic figure of dimensions used in Eq. \ref{eq:simplepower_appendix} }%
 \end{center}
 \end{figure}

The volumetric radiant emission power can be obtained by integrating the radiant intensity over a 4$\pi$ space angle, i.e.
\begin{equation}
\label{eq:RadiantEmissionPower_finite}
\frac{\rmd \Phi_e}{\rmd V}=4\pi\frac{\Delta\Phi_r}{\Delta V\Delta\Omega}
\end{equation}
By determining that $\Delta\Phi_r\equiv\Phi_r$, assuming that emission occurs in the middle of the plasma chamber ($X=D+\frac{L}{2}$) and combining Eqs. \ref{eq:finite_space_angle}, \ref{eq:LineOfSightVolume} and \ref{eq:RadiantEmissionPower_finite}, the volumetric radiant emission power can be expressed as
\begin{equation}
\label{eq:simplepower_appendix}
\frac{\rmd \Phi_e}{\rmd V}=\frac{4 \pi (D+\frac{L}{2})^2}{A_d A_e L(1+\frac{L}{D}+\frac{1}{3}\frac{L^2}{D^2})}\Phi_r.
\end{equation}
Equation \ref{eq:simplepower_appendix} is accurate, when detector is far away from the emission volume ($D\gg L$) and the aperture of the detector is small in comparison to the extraction aperture (Fig. \ref{fig:schematic}). 

\section*{References}

\bibliography{VUV_irradiance}

\end{document}